# Experiencing Extreme Height for The First Time: The Influence of Height, Self-Judgment of Fear and a Moving Structural Beam on the Heart Rate and Postural Sway During the Quiet Stance

**M. Habibnezhad[a], J. Puckett[a], M.S. Fardhosseini[b], H. Jebelli[c], T. Stentz[a], L.A Pratama[b]**

[a]Durham School of Architectural Engineering and Construction, University of Nebraska-Lincoln, Lincoln, NE, USA,68588
[b]Department of Construction Management, College of Built Environments, University of Washington, Seattle, WA, USA, 98195
[c]Department of Architectural Engineering, The Pennsylvania State University, 104 Engineering Unit A, University Park, PA,USA, 16802.2
E-mail: mahmoud@huskers.unl.edu , jay.puckett@unl.edu, sadrafh@uw.edu, jebelli.houtan@gmail.com, tstentz1@unl.edu, pragungl@uw.edu

**Abstract –**
   **Falling from elevated surfaces is the main cause of death and injury at construction sites. Based on the Bureau of Labor Statistics (BLS) reports, an average of nearly three workers per day suffer fatal injuries from falling. Studies show that postural instability is the foremost cause of this disproportional falling rate. To study what affects the postural stability of construction workers, we conducted a series of experiments in the virtual reality (VR). Twelve healthy adults—all students at the University of Nebraska-Lincoln—were recruited for this study. During each trial, participants' heart rates and postural sways were measured as the dependent factors. The independent factors included a moving structural beam (MB) coming directly at the participants, the presence of VR, height, the participants' self-judgment of fear, and their level of acrophobia. The former was designed in an attempt to simulate some part of the steel erection procedure, which is one of the key tasks of ironworkers. The results of this study indicate that height increase the postural sway. Self-judged fear significantly was found to decrease postural sway, more specifically the normalized total excursion of the center of pressure (TE), both in the presence and absence of height. Also, participants' heart rates significantly increase once they are confronted by a moving beam in the virtual environment (VE), even though they are informed that the beam will not 'hit' them. The findings of this study can be useful for training novice ironworkers that will be subjected to height and/or steel erection for the first time.**

## 1 Introduction

The growing rate of injuries and fatalities due to falling is concerning. In 2017, more than 5147 workers died at job sites, from which 887 were fall-related accidents [1]. Unfortunately, these fall-related losses have reached to the highest level in the 26-year history of the Census of Fatal Occupational Injuries (CFOI)[2]. This fact indicates that current safety procedures, training, and precautions have not adequately limited these types of injuries and fatalities at construction sites.

Therefore, numerous research studies have been conducted to shed lights on the important predictors of falling [3]. Some studies presented risk perception of construction workers as the key driver for these incidents, suggesting that the workers' engagement in unsafe situations results in severe injuries and death [4–6]. These types of studies are valuable in investigating the risk-taking behavior of construction workers [7]. However, the factors affecting postural stability of the workers can address more fall incidents, considering scenarios in which the worker has already executed the task [8].

Postural stability is the result of an association between three main sensory cues, namely, visual, vestibular, and somatosensory systems. The foremost mission of these balance control systems is to maintain the equilibrium of the center of mass of the body against direct or indirect stimuli. Accordingly, to find the root causes of postural instability, first, one needs to discover these provoking factors. Interestingly, from the three sensory cues for postural stability regulation, visual input is the proactive mechanism of balance, while the other two balance mechanisms of the body, triggered after the external exposure, considered reactive mechanisms. Therefore, the focus of this study would be mainly on the visual inputs as the dominant instigating factors. According to Hsiao and Simeonov, elevation, moving visual scenes, and depth perception are the most commonly visual perturbations affecting balance [8]. Not only these visual perturbations cause instigating mismatch and affect balance, but the exertion of anxiety due to the unsafe nature of these factors, especially those present at construction sites, will influence postural stability as well.

Following this rational, many studies strived to find out the extent to which anxiety (fear) impact postural regulation parameters. Interestingly, the result of these studies demonstrated that the presence of fear, especially in great extents, will negatively impact postural stability [9,10]. Notably, most of these research studies examined fear by utilizing various fear and acrophobia questionnaires. While these 'passive' questionnaires are reliable in measuring the self-judged level of anxiety, however, the need for more 'active' methods in determining the anxiety level of construction workers, subjected to fearful visual perturbations, seems necessary. Methods such as measuring variations in the facial temperature [11] the level of salivary cortisol [12], and heart rate [13] can imply the anxiety level. The latter could be misleading since many activities will impact the heart rate and diminish the reliability of the results. On the other hand, measuring heart rate is simple and inexpensive due to the ubiquity of smartwatches such as Fitbit and Apple watch. Considering the caveats above, this study attempts to investigate if heartrate variability across different trials is significantly affected by provoking visual factors.

## 2  Point of Departure

The current study will reinvestigate the impact of visual stimuli and fear on the postural stability through the precise and realistic simulation of the environment and virtual body parts (mixed reality). In other words, this research strives to create a virtual environment (VE) of the construction sites, in which participants will be exposed to 1. extreme height and 2. a moving structural beam directed towards participants as a part of the steel erection simulation. Also, the impact of these stimuli on the heart rate of the participant will be measured precisely. To our knowledge, very few studies utilize heart rate for these types of tasks happening at the construction sites.

## 3  Background

### 3.1  Impact of height on the visual sensory system.
As previously described, one of the most prominent reasons behind slip, trip, and falling is the loss of balance. As such, finding the influential factors which trigger instability in humans, especially those working at great height, is of utmost importance. The integration of the visual, vestibular, and somatosensory sensory systems is essential in maintaining one's postural balance. Some studies suggest that visual sensory system is the dominant system in controlling the human body's postural balance [14] and essential for maintaining balance [15–17]. In addition, the visual sensory system can compensate for the lack of other balance-dependent sensory systems, vestibular and somatosensory, in the presence of visually-induced disturbances. While each sensory system's weight on the postural stability is dependent on the environmental context and goal of the study [16], the importance of visual input as the prominent indicator of postural balance alludes that study of visual perturbations should be researched more extensively. Considering one in three people has visual height intolerance [18], many studies suggest height as an important stimuli in provoking gait instability and susceptibility to falling [19–22]. Studies examining height and fear of falling have reported that the deficit of close visual contact due to the instigating sensory mismatch, and fear-related reactions, especially close to edges of elevated surfaces, are the two main characteristics of falling from an elevated surface [8,23]. Following these findings, the current study aims to explore the aforementioned instigating factors, height, and the corresponding fear, on the postural sway of the participants subjected to virtual height.

### 3.2  Fear of height.
Fear is a bodily response to a danger or hazard in the form of emotional or physical reacts. Sometimes, these responses can be alarming, adversely influencing the humans' health and performance ranging from the reluctance of going to a dentist to changes in the gait and postural stability on elevated surfaces [24,25]. In light of the foregoing, it is important to consider fear and the fear of height (person's level of acrophobia) as leading factors influencing gait and postural stability [26]. Accordingly, in the current study, two questionnaires were presented to the participants, one for generally measuring fear and one for exclusively measuring the level of acrophobia of each subject. Although the focus of this study is on the height-related factors impacting safety, the circular contribution of visuo-vestibular to the fear and anxiety should not be ignored [21]. Coelho and Balaban stated that visuo-vestibular are not only associated with the fear of height, but also with panic and driving. Therefore, along with the acrophobia questionnaire (AQ), the famous James Geer fear measurement scale [27] has been utilized to pinpoint the potential relationship between any of the aforementioned fear aspects to the changes in the gait and postural stability parameters. Later, it will be shown that those questions related to sharp objects, driving, and auto accident, height, airplane, roller coasters, and death, appropriately predict height while only-height related questions were incapable of addressing postural parameter changes due to height. While these self-judgmental scales possibly show the self-perceived fear, however, the consideration of physiological responses to the experiment trials can be beneficial as well. The number of heartbeats per minute (HBM) is suggested to be a good indicator of the emotional responses to the unexpected thread or danger [13]. Therefore, in this study, the HBM factor was considered to be informative and included in the data collection process.

## 4 Methodology

**4.1 Participants.** Based on the flyer approved by the institutional review board (IRB), we recruited 13 students from the University of Nebraska-Lincoln (UNL). Although randomization was considered in the recruiting process, we attempted to selectively recruit an equal number of males and females, so that the potential influential sex factor could be precisely considered. As a result, six females and seven males were selected for of this study. The average age of the volunteers was 29 years old. The participants had no previous falling experience or any problem with standing still.

**4.2 Apparatuses and software.** To measure the center of pressure (COP) of the participants and consequently calculate the postural sway parameters, the AMTI force plate was utilized, which is capable of accurately calculating COP based on the reaction forces of the plate. Later, by utilizing an in-house code in MATLAB, all the necessary parameters for the postural sway, and heart rate would be imported and calculated. The HTC Vive Pro headset was selected for the immersive virtual environment demonstration. We chose the Vive Pro over other brands since the resolution and display quality of the headset is believed to be higher. As for the game engine software, the Unity3D software [28] was used, so that the creation of the immersive environment and all the VR simulations such as induction of height and the moving structural beam would be easily performed. Finally, to collect the heart rate of the participants, we used a Fitbit Versa. To easily and precisely synchronize the heart rate collection with the other devices, a heart rate collector application was developed in Fitbit studio, so that the start and stop events for each trial could be triggered on the Fitbit Versa remotely from a smartphone.

**4.3 Questionnaires.** Prior to the experiments, to predict the participants' self-judgment fear scale, the electronic version of James Geer's questionnaire [27] was presented to each participant. In addition, to measure their level of acrophobia, participants were asked to fill out the Cohen acrophobia questionnaire (AQ) as well [29]. As mentioned before, we attempted to find out if the results of these questionnaires can be promising in predicting the physiological responses such as heart rate or postural sway, in different study setups.

**4.4 Experiments.** The first part of the experiment was designed to help the participants get familiar with the VR environment. The VR environment was the same environment for all parts of the data collection. To potentially improve the feeling of presence in the VR environment, the VR model was enhanced with virtual legs. To ensure that the learning curve effect would be minimal for the participants, prior to conducting the experiment, they were asked to stand, walk, and look at their virtual legs for 1 minute. The force plate was placed on a specific location in an office so that the initial standing position of the participants became the designated location in the virtual environment (VE). In the second part of the experiment, the participants were asked to stand in the center of the force plate and open their legs to the extent to which they are comfortable and most stable. After 5 seconds, the participants were asked to look at their feet once, for 2 seconds, with the least movement possible. This movement will affect the postural stability of the participants; however, they were instructed to perform the same task in the other trials. Therefore, the data collection would be consistent and not biased. This trial finished after 20 seconds. During the trial, the heart rate and COP of the participants were retrieved. After the completion of the first part, the participants were equipped with the VR headset, while they were asked to hold the initial position of their feet on the force plate. In the second trial, the same procedure as the previous trial was undertaken, however, virtually.

The participants were asked to look at their 'virtual feet' in the same way as before. To ensure that the experiment complies with that of the quiet stance postural balance, no subject was allowed to look down more than once. After 20 seconds, a moving structural beam hung from a crane wire, slowly approached the participants. The moving beam's trajectory was in line to the participant's site of view. Before the start of this trial, each participant was informed that the virtual beam will not 'hit' them and will stop 1ft away from their body. This part of the experiment took 10 seconds to finish. The last part of the experiment was almost identical to the second part of the experiment, however, this time the participants were placed on the 17th floor of an unfinished building. The VE, standing beam, moving beam, lighting, shadows, and other virtual scenes were identical to the pervious trial, except the height. Again, before the start of this trial, all the participants were informed that the moving beam will slowly approach them and will not 'hit' or 'pass through' them. With the last 10 seconds of the moving structural beam approaching the participants, the overall duration of this part was 30 seconds.

Table 1 Experiment configuration

| Scenarios | | | | | |
|---|---|---|---|---|---|
| No Height | | | Height | | |
| No VR | VR - No MB | VR - MB | No VR | VR - No MB | VR - MB |

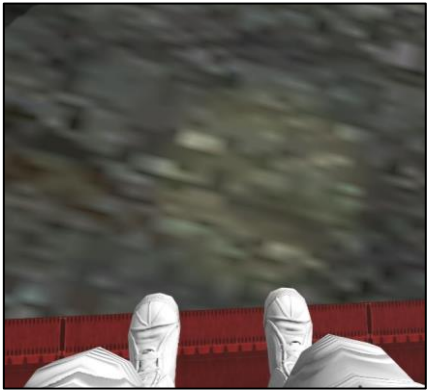

(a)

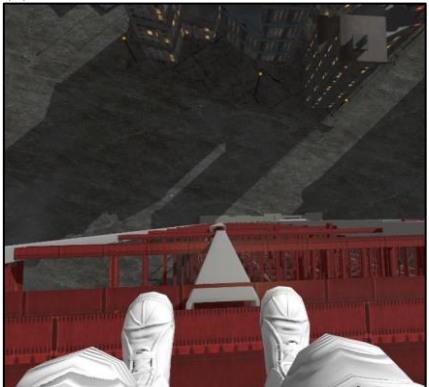

(b)

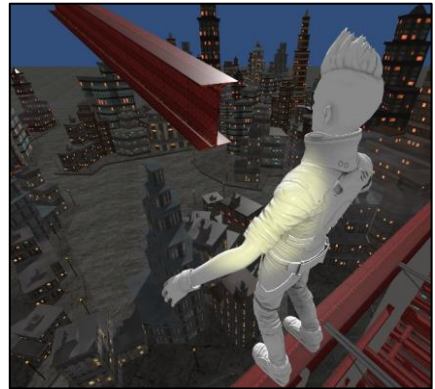

(c)

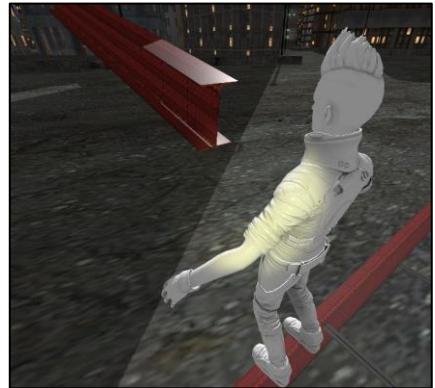

(d)

Figure 1. All the virtual scenarios in the experiments. The (a) image is no height and no MB scenario, and the (b) is height and no MB. The (c) image is no height and MB scenario, and the (d) image is height and MB.

## 5 Results and Analysis

All the data was imported from different text files generated throughout the experiments and then tabulated for statistical analysis. By using the MATLAB statistical toolbox, standard two-tail T-test was performed on the table of experiment results. It is important to note that the normality of the data was tested by performing the Anderson-Darling test [30] and turned out to be valid for this study. For the first part of the analysis, all the data was tabulated for a within-subject design case in which the same subjects perform in all the levels of independent variables. The result of the paired T-tests for such an analysis can be investigated in Table 2. The dependent factors were heart rate and postural parameters. The independent variables were the presence of VR, Height, MB on the ground (no height), and MB at height. Four postural parameters were considered in this research: 1. Total Excursion (TE) of the COP throughout the trial duration, 2. Root-mean-square (RMS) amplitude is representing the variability (standard deviation) of the COP displacement. 3. Maximum absolute amplitude distance of the COP traveled during the trial duration, and 4. Peak-to-Peak (PP) amplitude is expressing the maximum displacement shown between two COP points during the trial. As shown in Table 2, there was a statistically significant difference between the heart rate of the participants subject to the MB on the ground and at height with the P value of 0.00005. Based on the descriptive results, the HRMs of the participants increased by almost 10 percent. Another interesting case was the significant increase in the TE of the participants exposed to VR for the first time. Although prior to the data collection, they experienced VR, the significant increases in all the postural parameters were noticeable.

Table 2 T test's significant level of mean differences between each pair of groups (P values)

|  | P-values | | | |
|---|---|---|---|---|
|  | Presence of VR | Height | MB (no Height) | MB (Height) |
| Heart rate | .190 | .554 | .00005** | .00005** |
| Postural sway (TE) | .004** | .052* | .497 | .890 |
| Postural sway (RMS) | .066 | .174 | .095 | .584 |
| Postural sway (Max) | .057 | .156 | .112 | .694 |
| Postural sway (PP) | .017* | .084 | .123 | .670 |

\* Sig. at 0.05 level
\*\* Sig. at 0.01 level

Table 3 Descriptive standard two-tail unpaired T-test mean results for HR and RMS based on the participants' sex, fear (selective questions) and AQ

|  |  | No height | | | Height | |
|---|---|---|---|---|---|---|
|  | Groups | No VR | VR - No MB | VR – MB | VR - No MB | VR – MB |
| Sex (HRM) | Male | 83.23 | 82.31* | 90.65* | 80.74* | 89.65 |
|  | Female | 86.73 | 99.18* | 111.42* | 98.26* | 105.98 |
| Fear[1] (HRM) | Low[2] | 85.98 | 87.75 | 98.73 | 87.83 | 98.11 |
|  | High | 83.98 | 90.93 | 99.88 | 88.25 | 97.51 |
| AQ (HRM) | Low | 85.06 | 88.01 | 98.18 | 89.43 | 94.95 |
|  | High | 84.90 | 90.66 | 100.43 | 86.65 | 100.68 |
| Sex (RMS) | Female | .000217 | .000341 | .000350* | .000447 | .000265* |
|  | Male | .000284 | .000530 | .001318* | .000588 | .000845* |
| Fear (RMS) | Low | .000249 | .000292 | .000693 | .000275* | .000395 |
|  | High | .000251 | .000579 | .001136 | .000784* | .000812 |
| AQ (RMS) | Low | .000199 | .000370 | .000591 | .000336 | .000341 |
|  | High | .000302 | .000501 | .001238 | .000722 | .000865 |

Besides, Table 2 demonstrates that the mean difference between the TE of the subjects collected during the quiet stance in VR on the ground and their TE during the quiet stance in VR at height was statistically significant with the P value of 0.05. In other words, the TE of the participants increased when they were subjected to height. For each trial, to measure the impact of the participants' self-judged fear values on the heart rate and postural stability, an in-between subject design was performed on the tabulated data. James Geer's fear questionnaire consists of 50 questions collected based on the empirical data and participants' reports [27]. The questions range from auto accidents, being a passenger in an airplane, and height to hypodermic needles, being alone, and crowded places. Since many of these questions may not be strongly correlated to our physiological responses at height, we conducted various test aimed to find out which of these questions best describe the fear of height. Considering Coelho and Balaban's visuo-vestibular study [21], and the result of this study, we find out that "sharp edges", "being a passenger in an airplane", "roller coasters", "death", "heights", "driving a car", and "auto accidents", are strongly in line with the visuo-vestibular predicting factors and leads to statistically significant results in our study. To account for the factor of sex, the participants were divided into two groups of male and female, respectively. The data was then sorted based on the level of fear reported by the subjects and split into two groups of less fearful and more fearful subjects. The same procedure was undertaken for the AQ factor. The result of the unpaired T-tests on the two groups of the participants can be seen in table 3.

By focusing on the P values of the T-tests on each of the paired groups, the significant mean differences between the groups are recorded in table 3. As reported in this table, in VR with no height and the presence of MB, the male group has a statistically significant lower average HRM compared to the female group. Interestingly, the difference between the RMS of the 'fearful' and 'non-fearful' group was significant (P value = 0.023) for the same scenario, VR with MB and no height. Besides, the presence of height in VR played a key role in differentiating between the RMS of the fearful and non-fearful participants; The participants with more self-reported fear had a higher RMS mean compared to those with less self-reported fear (P-value = 0.046). A similar effect was manifested when the two male and female groups were compared concerning their HR differences during the quiet stance at height. Females had higher HR compared to males. Notably, no statistical difference was spotted between the male and female groups for RMS at height in the absence of MB. However, in the presence of MB at height, the result of T-test indicated that the mean RMS of the females was significantly higher (P value = 0.009) compared to that of males. Another important observation was the inability of AQ values in predicting the differences between the two groups of subjects separated by their levels of acrophobia.

## 6    Discussion and Conclusion

According to the statistical test results, the presence of a virtual environment has a substantial influence on the postural sway of the participants. Except for postural variability, RMS, almost all the other postural parameters increased for the participants who were experiencing VR for the first time. Based on similar findings of previous studies, the observed increase in the postural sway—due to the exposure to the immersive environment—was predictable [31].

Furthermore, participants were asked to 'normally' look at their feet during the quiet stance in VR with the absence of height. Due to the lower field of view of the VR headsets compared to that of a person without wearing VR headsets, participants' postural balance might have been affected. Streepey et al., pointed out this limitation as a lack of peripheral visual inputs while wearing glasses [32]. Therefore, although the existence of such a factor in the comparison between VR trial cases is unlikely, its potential impact on the presence of VR cannot be rendered negligible. As for the impact of elevated surfaces on balance, the presence of height increased subjects' postural sway. Based on the statistically significant mean differences of the participants' TE, the presence of height factor was able to predict the difference in the means with the P value of 0.05. This finding was also in line with the results of the Cleworth et al., study [33]. They suggested that the virtual height reduces the mediolateral and anterior-posterior balance of subjects similar to the real height. Remarkably, the visual stimuli in the form of a moving object increased participants heartbeat dramatically. With a P value of 0.00005 both with and without the presence of height, the MB significantly increased participants' heart rates. These findings suggest that tasks involving handling of big moving objects, e.g., steel erection, will increase the heart rate of first-time workers exposed to the great extents. This physiological response to big approaching objects should be studied more extensively since the noticeably low P-value magnifies its importance in the context of safety. On the other hand, since the participants were informed that the approaching digital beam would not 'hit' them, no statistically significant mean changes were observed for their postural parameters.

Based on the results of the unpaired T-tests, those fear questions related to driving, height, and sharp objects, derived from James Geer's fear questionnaire, powerfully predict the differences in the RMS in the presence of virtual height. Accordingly, it can be inferred that fear increases RMS, in the presence of virtual height. In other words, standing on elevated surfaces, people with higher self-judged fear values have more postural sway variability compared to those with lower fear values. Adkin et al. showed that fear of height modifies the postural sway [34]. Based on the participants' anxiety rating, they suggested that balance would be affected by the fear of height factor. Similar to our findings, they concluded that both physiological variables, such as postural balance, and psychological factors, such as anxiety, may result in postural instability.

However, to our surprise, the AQ values were not able to address any changes in the postural stability, nor were they able to explain heart rate for different scenarios during quiet stance. One explanation for the incapability of AQ to address mean differences between the corresponding groups might be because this questionnaire was designed to detect abnormalities rather than differences concerning the fear of height. Therefore, since most of the participants did not score high on this questionnaire, then they latent differences regarding fear of height was not noticeable. Following this rationale, a threshold-based categorization should be conducted on the AQ results rather than the AQ mean-based approach. Since the threshold should not be less than 2 (Likert-scale range for AQ is from 1 to 5), so that it can detect people with acrophobia, and most of our subjects' average AQ score was less than 2, we were not able to categorize with thresholds.

With respect to subjects' gender, at height, female subjects perceived a higher level of anxiety, in the form of HR, compared to male. Also, the average HR of female participants were higher compared to that of male, when they were confronted by the moving structural beam. Another interesting result was the influence of gender on the RMS. With the P values less than 0.05, their RMS values were significantly lower than male in both scenarios with MB. Therefore, regarding bodily responses in the proximity of big moving objects, seemingly, women HR increases more than men and men's RMS increases more than women. Considered as one of the limitations of the current study, prior information regarding VR studies could influence the postural stability of the participants as their uncertainty about the experiments mixes with the expectation of being at height. This unwanted effect potentially explains the differences between the postural stability of male and female groups in VR with the absence of any visual perturbations.

## References

[1]   Bureau of Labor Statistics Data, (2018). https://data.bls.gov/timeseries/FWU00X4XXX XX8EN00 (accessed January 10, 2019).

[2]   NATIONAL CENSUS OF FATAL OCCUPATIONAL INJURIES IN 2017, (2018). www.bls.gov/iif/oshcfoi1.htm (accessed July 9, 2018).

[3]   M.S. Fardhosseini, B. Esmaeili, The Impact of the Legalization of Recreational Marijuana on Construction Safety, in: Constr. Res. Congr. 2016, American Society of Civil Engineers,

Reston, VA, 2016: pp. 2972–2983. doi:10.1061/9780784479827.296.

[4]   T. Abdelhamid, J. Everett, Identify Root Causes of Construction Accidents, J. Constr. Eng. Manag. 126 (2000) 52–60. doi:10.1061/(ASCE)0733-9364(2000)126:1(52).

[5]   H. Mahmoud, F. Sadra, V.A. Moghaddam, E. Behzad, D.M. D., The Relationship between Construction Workers' Risk Perception and Eye Movement in Hazard Identification, in: Constr. Res. Congr. 2016, 2017. doi:doi:10.1061/9780784479827.297.

[6]   S. Boheir, S. Hasanzadeh, B. Esmaeili, M.D. Dodd, S. Fardhosseini, MEASURING CONSTRUCTION WORKERS' ATTENTION USING EYETRACKING TECHNOLOGY, in: 5th Int. Constr. Spec. Conf., Univ. of British Columbia, Vancouver, 2015. https://www.researchgate.net/publication/28851 6371 (accessed March 21, 2019).

[7]   M. Habibnezhad, B. Esmaeili, The Influence of Individual Cultural Values on Construction Workers' Risk Perception, in: 52nd ASC Annual International Conference Proceedings, 2016. http://ascpro0.ascweb.org/archives/cd/2016/pap er/CPRT211002016.pdf (accessed October 30, 2017).

[8]   H. Hsiao, P. Simeonov, Preventing falls from roofs: a critical review, Ergonomics. 44 (2001) 537–561. doi:10.1080/00140130110034480.

[9]   C.C. Boffino, C.S. Cardoso de Sa, C. Gorenstein, R.G. Brown, L.F.H. Basile, R.T. Ramos, Fear of heights: Cognitive performance and postural control, Eur. Arch. Psychiatry Clin. Neurosci. 259 (2009) 114–119. doi:10.1007/s00406-008-0843-6.

[10]  H.T. Regenbrecht, T.W. Schubert, F. Friedmann, T.W. Schubert Ifrank Friedmann, Measuring the Sense of Presence and its Relations to Fear of Heights in Virtual Environments, Int. J. Hum. Comput. Interact. 10 (1998) 233–249. doi:10.1207/s15327590ijhc1003_2.

[11]  I. Pavlidis, J. a Levine, P. Baukol, Ioannis Pavlidis Honeywell Laboratories Minneapolis , MN ioannis . pavlidis @ i ), honeywell . com James Levine Mayo Clinic Mayo Clinic, IEEE Trans. Biomed. Eng. (2002) 315–318.

[12]  S.A. Vreeburg, F.G. Zitman, J. Van Pelt, R.H. Derijk, J.C.M. Verhagen, R. Van Dyck, W.J.G. Hoogendijk, J.H. Smit, B.W.J.H. Penninx, Salivary cortisol levels in persons with and

without different anxiety disorders, Psychosom. Med. 72 (2010) 340–347. doi:10.1097/PSY.0b013e3181d2f0c8.

[13] C. Schmitz, L. Drake, M. Laake, P. Yin, R. Pradarelli, Physiological Response to Fear in Expected and Unexpected Situations on Heart Rate, Respiration Rate and Horizontal Eye Movements, J. Adv. Student Sci. 1 (2012). http://jass.neuro.wisc.edu/2012/01/Lab 602 Group 10 Final Submission.pdf (accessed December 31, 2018).

[14] M.G. Gaerlan, The role of visual, vestibular, and somatosensory systems in postural balance, 2010. http://digitalscholarship.unlv.edu/thesesdissertat ions (accessed December 31, 2018).

[15] M. Friedrich, H.J. Grein, C. Wicher, J. Schuetze, A. Mueller, A. Lauenroth, K. Hottenrott, R. Schwesig, Influence of pathologic and simulated visual dysfunctions on the postural system, Exp. Brain Res. 186 (2008) 305–314. doi:10.1007/s00221-007-1233-4.

[16] L.M. Nashner, C.L. Shupert, F.B. Horak, F.O. Black, Organization of posture controls: An analysis of sensory and mechanical constraints, Prog. Brain Res. 80 (1989) 411–418. doi:10.1016/S0079-6123(08)62237-2.

[17] M. Schmid, A. Nardone, A.M. De Nunzio, M. Schmid, M. Schieppati, Equilibrium during static and dynamic tasks in blind subjects: No evidence of cross-modal plasticity, Brain. 130 (2007) 2097–2107. doi:10.1093/brain/awm157.

[18] D. Huppert, E. Grill, T. Brandt, Down on heights? One in three has visual height intolerance, J. Neurol. 260 (2013) 597–604. doi:10.1007/s00415-012-6685-1.

[19] F. Ayoubi, C.P. Launay, A. Kabeshova, B. Fantino, C. Annweiler, O. Beauchet, The influence of fear of falling on gait variability: results from a large elderly population-based cross-sectional study., J. Neuroeng. Rehabil. 11 (2014) 128. doi:10.1186/1743-0003-11-128.

[20] M.E. Chamberlin, B.D. Fulwider, S.L. Sanders, J.M. Medeiros, Does Fear of Falling Influence Spatial and Temporal Gait Parameters in Elderly Persons Beyond Changes Associated With Normal Aging?, 2005. https://core.ac.uk/download/pdf/85214868.pdf (accessed December 30, 2018).

[21] C.M. Coelho, C.D. Balaban, Visuo-vestibular contributions to anxiety and fear, Neurosci. Biobehav. Rev. 48 (2015) 148–159.

doi:10.1016/J.NEUBIOREV.2014.10.023.

[22] P.I. Simeonov, H. Hsiao, B.W. DotsonM, D.E. Ammons, HEIGHT IN REAL AND VIRTUAL ENVIRONMENTS, Hum. Factors J. Hum. Factors Ergon. Soc. 6 (2005) 430–438. https://journals.sagepub.com/doi/pdf/10.1518/00 18720054679506 (accessed December 31, 2018).

[23] T. Brandt, F. Arnold, W. Bles, T.S. Kapteyn, The mechanism of physiological height vertigo. I. Theoretical approach and psychophysics., Acta Otolaryngol. 89 (1980) 513–23.

[24] J.R. Davis, A.D. Campbell, A.L. Adkin, M.G. Carpenter, The relationship between fear of falling and human postural control, Gait Posture. 29 (2009) 275–279. doi:10.1016/J.GAITPOST.2008.09.006.

[25] S.M. Gordon, Dental fear and anxiety as a barrier to accessing oral health care among patients with special health care needs, Spec. Care Dent. 18 (1998) 88–92. doi:10.1111/j.1754-4505.1998.tb00910.x.

[26] C.M. Coelho, G. Wallis, Deconstructing acrophobia: physiological and psychological precursors to developing a fear of heights, Depress. Anxiety. 27 (2010) 864–870. doi:10.1002/da.20698.

[27] J.H. Geer, The Development Of A Scale To Measure Fear, Behav. Res. Ther. 3 (1965) 45–53.

[28] Unity - Game Engine, (2017). https://unity3d.com/ (accessed July 29, 2017).

[29] D.C. Cohen, Comparison of self-report and overt-behavioral procedures for assessing acrophobia, Behav. Ther. 8 (1977) 17–23. doi:10.1016/S0005-7894(77)80116-0.

[30] T.W. Anderson, D.A. Darling, Asymptotic Theory of Certain "Goodness of Fit" Criteria Based on Stochastic Processes, Ann. Math. Stat. 23 (1952) 193–212. doi:10.1214/aoms/1177729437.

[31] C.G.C. Horlings, M.G. Carpenter, U.M. Küng, F. Honegger, B. Wiederhold, J.H.J. Allum, U.M. Küng, F. Honegger, B. Wiederhold, J.H.J. Allum, Influence of virtual reality on postural stability during movements of quiet stance, Neurosci. Lett. 451 (2009) 227–31. doi:10.1016/j.neulet.2008.12.057.

[32] J.W. Streepey, R. V. Kenyon, E.A. Keshner, Field of view and base of support width influence postural responses to visual stimuli during quiet stance, Gait Posture. 25 (2007) 49–55.

doi:10.1016/j.gaitpost.2005.12.013.

[33]    T.W. Cleworth, B.C. Horslen, M.G. Carpenter, Influence of real and virtual heights on standing balance, Gait Posture. 36 (2012) 172–176. doi:10.1016/j.gaitpost.2012.02.010.

[34]    A.L. Adkin, J.S. Frank, M.G. Carpenter, G.W. Peysar, Fear of falling modifies anticipatory postural control, Exp. Brain Res. 143 (2002) 160–170. doi:10.1007/s00221-001-0974-8.